\documentclass[12pt]{article}
\usepackage[utf8]{inputenc}
\usepackage{authblk}
\usepackage{sectsty}
\usepackage{hyperref}
\usepackage{graphicx}
\usepackage{booktabs}
\usepackage{amssymb}
\usepackage{amsmath}
\usepackage{amsmath,textcomp}
\usepackage{indentfirst}
\usepackage{xcolor}
\usepackage{xcolor}
\usepackage[left=2cm,right=2cm,top=1.5cm,bottom=3cm]{geometry}
\definecolor{light-gray}{gray}{0.95}

\title{\textbf{The law of momentum conservation in free microwave cavities containing coherent radiation}}
\author{Ademir Xavier Jr.\footnote{Presently with the Brazilian Space Agency, Brasília - DF, Brazil. E-mail: \textit{xavnet2@gmail.com}.}}
\date{\today}

\begin{document}
\maketitle
\begin{abstract}
Using classical electrodynamics, this work analyzes the dynamics of a closed microwave cavity as a function of its center of energy. Starting from the principle of momentum conservation, expressions for the maximum electromagnetic momentum stored in a free microwave cavity are obtained. Next, it is shown that, for coherent fields and special shape conditions, this momentum component may not completely average out to zero  when the fields change in the transient regime. Non-zero conditions are illustrated for the asymmetric conical frustum whose exact modes can not be calculated analytically. One concludes that the electromagnetic momentum can be imparted to the mechanical body so as to displace it in relation to the original center of energy. However, the average time range of the effect is much shorter than any time regime of the experimental tests performed to measure presently, suggesting it has not been observed yet in copper-made resonators. 
    
Keywords: microwave cavities, resonators, electromagnetic propulsion, electrodynamics, thrust anomaly.
\end{abstract}

\section{Introduction}

Electromagnetic (EM) radiation is an effective propulsion mechanism, despite the small thrust demonstrated in the laboratory. Fundamentally, the principle does not differ from the one of any chemical rocket, except for the utilization of photons as massless {ejecta} \cite{sherburne1953momentum,  wickman1981technology, bae2012prospective}.  The photon source may be powered by energy collected from the outside \cite{wickman1981technology} (e. g., a nearby star), dispensing, on the whole, the transportation of exhaustible fuel tanks. On the downside, practical accelerations provided by even the most powerful laser beams are so tiny \cite{marx1966interstellar} that handy payloads would take hundreds of years to achieve a fraction of the velocity of light.

There have been some recent examples of EM systems claiming to generate propulsion as reactionless thrusters \cite{white2017, duif2017improved, kossling2019spacedrive}, that is, without any ejection mechanism. These works were born as experimental investigations after initial non-zero thrust measurements in a certain kind of truncated cavity \cite{mullins2006fly, shawyer2015second} powered by microwave radiation (later known as ``EMdrive'').  Despite the numerous attempts to explain the effect \cite{juan2013prediction, mcculloh2017, fetta2014numerical, grahn2016exhaust}, no convincing theory was accepted to solve the apparent anomaly, and the interest in such devices faded away to some extent. Recently, it received a serious blow after sensitive experimental tests \cite{tajmarspace}, suggesting all previous results were caused by mechanical stresses due to thermal dilatation of the cavity body. Despite this negative report, a thorough theoretical account of the role played by momentum conservation in such system is still missing because electromagnetic fields also carry momentum which is conserved for the whole system only.

Given the extraordinary success of the EM theory, the problem is analysed here in accordance with the principle of momentum conservation, by realizing the device is in fact a coupled system of mechanical and electromagnetic subsystems, and by exploring the role played  by both the momentum stored in the EM fields and the center of energy (CE) \cite{einstein1906prinzip}. This also means that the possible electromechanical cause is singled out by neglecting important effects  competing in magnitude with the electrodynamic forces. In this sense, this study admits an ideal cavity. For example, by considering the cavity body as made of a material with low thermal dilatation coefficient, so that the produced RF heat will not affect the mass distribution about the original center of mass (CM). It is clear that the experimental characterization of the putative electromechanical effect described here is difficult, given the competing effects and the minute forces involved. The author hopes this work highlights the electromagnetic variables that may contribute the most to any measurable behaviour in addition to thermal effects or other traceable perturbations which are clearly present in the real device.


This work is organized as follows:  Section \ref{conservationlaws} presents a brief review of the conservation laws involving matter, charges and EM fields. These principles are then applied to a free microwave cavities in Section 3. Instead of calculating internal forces, this work focus on finding the time evolution of the partial EM momentum arising in the hollow space when its generator is turned on or off. This is done by deducing a relation between the EM and the body mechanical momenta after electromagnetic fields are set up within the cavity (Section 3.1). The conditions for non-zero displacement when a single mode is excited is analysed for both time (Section 4.1) and space (Section 4.3) averages. Relations for the total electromagnetic momentum stored in the cavity are obtained in Section \ref{estimatingmomentum}. Expressions for the internal thrust due to momentum variations in the transient phase are given in Section \ref{momentumvariation}. Finally, the conclusion is presented in Section \ref{conclusions}. 

\section{Conservation laws}
\label{conservationlaws}

It is well known that Newton's third law \textit{is not obeyed} by the pairwise magnetic force exchanged by two currents or moving charges if the momentum stored in their accompanying fields is not taken into account \cite{keller1942newton, zangwill2013modern}. For example, in the motion of a system of two free charged particles, the total mechanical momentum will be observed to change even when no external force exists. The paradox is solved by including an integral of the field momentum density generated by the set of particles. Such fundamental role played by EM fields in momentum conservation has given rise to the controversies about the ``hidden momentum'' as described in previous works \cite{keller1942newton, coleman1968origin, calkin1971linear, pugh1967physical, graham1980observation}. However, such studies have considered only static fields while the case at hand involves alternating fields. 

To account for the contribution of EM momentum, a system in vacuum and composed of particles and EM radiation has its total momentum $\vec{P}$ given in term of a sum of two terms \cite{stratton2007electromagnetic, zangwill2013modern, boyer2005illustrations}
\begin{equation}\label{eq01}
\vec{P}=\sum_i \gamma_i m_i \vec{v}_i+\int  \vec{g}(\vec{r}\,',t)d^3r', 
\end{equation}
with $m_i$ the mass of i-th particle composing the body,  $\gamma_i=(1-v_i^2/c_0^2)^{-1/2}$,  $\vec{v}_i$ the particle velocities, $c_0=\sqrt{1/\epsilon_0\mu_0}$ the speed of light in free space, and $\epsilon_0$ and $\mu_0$ the electric permittivity and magnetic permeability of free space respectively. Here $\vec{g}(\vec{r}\,',t)$ is Minkonwski's version of the density of EM momentum \cite{zangwill2013modern, stratton2007electromagnetic} obtained by vector multiplying the electric displacement $\vec{{D}}(\vec{r}\,',t)$ and magnetic induction $\vec{{B}}(\vec{r}\,',t)$ produced by the set of particles at point $\vec{r}\,'$ and time $t$ in the laboratory reference system.  The integral in Eq. (\ref{eq01}) using the volume element $d^3r'$ is over all space.  Conservation of momentum - and the validity of Newton's third law - implies that, in the absence of external forces, the total momentum $\vec{P}$ is constant and not only the mechanical momentum.

Two other integrals can be written for the system. Conservation of energy demands that the total energy \cite{coleman1968origin, boyer2005illustrations} 
\begin{equation}\label{eq02}
U=\sum_i\gamma_i m_i c_0^2 +\int  u(\vec{r}\,',t) d^3r'
\end{equation}
be constant as well, where
\begin{equation}\nonumber
u(\vec{r}\,',t)=\frac{1}{2}[ \vec{{E}}(\vec{r}\,',t)\cdot\vec{{D}}(\vec{r}\,',t)+\vec{{H}}(\vec{r}\,',t)\cdot\vec{{B}}(\vec{r}\,',t)]
\end{equation}
is the (scalar) energy density at  $\vec{r}$ and $t$, $\vec{{E}}(\vec{r}\,',t)$ and $\vec{{H}}(\vec{r}\,',t)$ are the electric and magnetic fields respectively. From Eq.(\ref{eq02}), the total system mass is given by $M=U/c_0^2$. Also, the CE vector $\vec{R}_{CE}$ of the system \cite{einstein1906prinzip, zangwill2013modern, moller1972theory, boyer2005illustrations,coleman1968origin}
\begin{equation}\label{eq03}
M\vec{R}_{CE}=\sum_i\vec{r}\,'_i\gamma_i m_i  +\frac{1}{c_0^2}\int  \vec{r}u(\vec{r}\,',t)d^3r' 
\end{equation}
is a constant of motion when no external forces are present. The first term on the right side (RHS) of Eq. (\ref{eq02}) includes all possible energy contributions intrinsic to the particles and that are not of electromagnetic origin. It is easily seen that, in the limit of $c_0\rightarrow\infty$, Eq.(\ref{eq03}) together with Eq.(\ref{eq02}) converge to the definition of the CM of the system \cite{boyer2005illustrations}, located at $\vec{R}_{CM}$.

\section{Application to a free microwave cavity}

\begin{figure}[ht]
\centering
  \includegraphics[width=75mm]{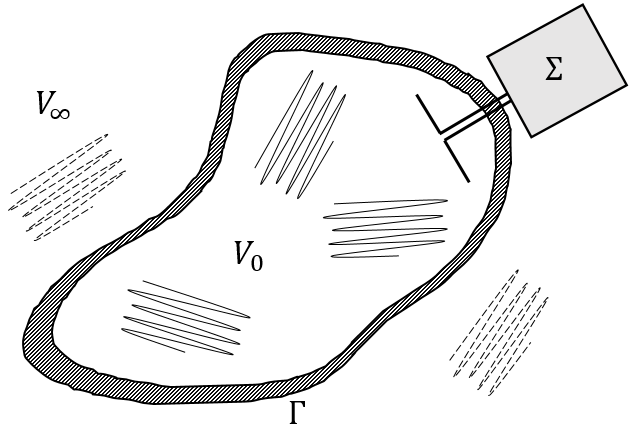}
  \caption{A free cavity of homogeneous material density $\rho$ carring an internal source or generator $\Sigma$ and containing electromagnetic fields in $V_0$ (hollow space) and $V_{\infty}$. In principle, unbound fields exists in $V_{\infty}$ outside the cavity boundary, depending on the thickness of $\Gamma$ and the generator power intensity.}
  \label{fig:01}
\end{figure}

In the spirit of \cite{penfield1966electrodynamics}, Figure \ref{fig:01} splits the cavity into two subsystems: the cavity body $\Gamma+\Sigma$ (with $\Sigma$ representing the generator) and the EM fields in $V_0$ and partially penetrating $\Gamma$. The cavity body is admitted as made of a homogeneous material of finite conductivity $\sigma$ and initial mass density $\rho$. The magnetic permeability and electric permittivity of both cavity and generator material are homogeneous, isotropic and equal to $\mu$ and $\epsilon$, respectively, so that $\vec{D}=\epsilon\vec{{E}}$ and $\vec{{B}}=\mu\vec{{H}}$.  The generator feeds  the internal volume $V_0$ with EM energy, and unbound fields may exist in the external space $V_{\infty}$. For small velocities, $\gamma(v)\approx 1$, and the total momentum is approximated as
\begin{equation}\label{eq4}
\vec{P}=\int_{\Gamma+\Sigma}  \rho'\vec{v}d^3r'+\frac{1}{c^2}\int_{\Gamma+\Sigma}  \vec{S}(\vec{r}\,',t)d^3r'+\frac{1}{c_0^2}\int_{V_0+V_{\infty}} \vec{S}(\vec{r}\,',t)d^3r',   
\end{equation}
with $\vec{S}(\vec{r}\,',t)=\vec{{E}}(\vec{r}\,',t)\times\vec{{H}}(\vec{r}\,',t)$, $c=(\mu\epsilon)^{-1/2}$ the speed of light in $\Gamma$ and $\Sigma$, and assuming for simplicity $V_0$ and $V_{\infty}$ as vacuum. In going from Eq. (\ref{eq01}) to Eq. (\ref{eq4}) one has assumed that all electromagnetic (unaffected) fields associated with atomic matter (``\textit{die ponderabele masse}'' according to \cite{einstein1906prinzip}) are represented by the term containing the continuous mass density distribution $\rho'$, and that in Eq. (\ref{eq5}) the vector $\vec{S}$ describes fields induced in $\Gamma$ by the generator only.

For typical microwave frequencies ($\omega/2\pi\approx 1$GHz), and cooper of conductivity $\sigma = 5.9\times 10^{7}$ S/m, the skin depth $\delta=(2/\omega\mu\sigma)^{1/2}$ is about 2 $\mu$m.  Therefore, for frequencies in the microwave range, the unbound integral can be neglected if the cavity is thick enough because the fields are evanescent within $\Gamma$ and $\Sigma$. One also assumes that $V_{\infty}$ does not contain any other fields produced by microwave radiation damping in $\Gamma$. 

Besides the restriction to confined fields, a crucial assumption is to consider the cavity as a rigid body which allow us to immediately identify the integral of the mechanical momentum as $M\vec{V}_{CM}$, with $\vec{V}_{CM}=d\vec{R}_{CM}/dt$, the velocity of the CM as defined by Eq. (\ref{eq03}) when all internal fields are zero. Such approximation is reasonable if the intensity of the radiation force on the internal parts of the cavity is much smaller than the typical mechanical stresses necessary to deform the structure. Moreover, the propagation time of EM radiation in $V_0$ is considered much shorter than propagation time of mechanical waves in $\Gamma$. When $t<t_0$, the generator is turned off. Hence, the cavity initial momentum is $\vec{P}(t<t_0)=M_0{\vec{V}}_{CM}$, with ${\vec{V}}_{CM}\ll c$, and 
\begin{equation}\nonumber
M_0=\int_{\Gamma+\Sigma} \rho d^3r'
\end{equation}
is the total mass. For $t\ge t_0$ the generator is turned on and the fields fill up the volume $V_0$ and penetrate $\Gamma$ and $\Sigma$, obeying well-known boundary conditions between free-space and the conductor. For $t\ge t_0$ and defining $\Delta \vec{v}=\vec{v}-{\vec{V}}_{CM}$, $\vec{r}=\vec{r'}-\vec{R}_{CM}$, Eq. (\ref{eq4}) may be rewritten as
\begin{equation}\label{eq5}
\begin{split}
M\Delta\vec{v}&=-\frac{1}{c^2}\int_{\Gamma+\Sigma}  \vec{S}(\vec{r},t)d^3r - \frac{1}{c_0^2}\int_{V_0}\vec{S}(\vec{r},t)d^3r\\&+ \left(\frac{1}{c^2}\int_{\Gamma+\Sigma}  u(\vec{r},t) d^3r+\frac{1}{c_0^2}\int_{V_0}u(\vec{r},t)d^3r\right)\vec{V}_{CM},   
\end{split}
\end{equation}
where 
\begin{equation}\nonumber
M=\int_{\Gamma+\Sigma} \rho' d^3r
\end{equation}
is the cavity mass excluding the fields, and the energy density $u(\vec{r},t)$ refers to the fields generated by the source only. This result implies in the possible onset of a non-zero velocity change of the cavity body in relation to $\vec{R}_{CM}$, whenever the EM field changes, as is the case when the generator is turned on or off. To avoid other complications (e. g. Doppler shift of the fields), one assumes the cavity is at rest in the laboratory frame for $t<t_0$, that is, $\vec{V}_{CM}=0$ making the second term on the RHS of Eq.(\ref{eq5}) equal to zero (this term is equivalent to the linear momentum associated with the ``rest mass'' of the fields).  Hence, possible field configurations resulting in non-zero values for the integrals in Eq. (\ref{eq5}) should be analysed with care.

\subsection{Condition for non-zero $\Delta \vec{v}$}

From now on, one takes the origin of the coordinate system as the initial CM, so that $\vec{R}_{CM}=0$. In principle, for a time dependence of the form $e^{i\omega t}$, $\langle \vec{E}\times\vec{{H}}\rangle \propto \vec{{E}}\times\vec{{H}}^{\star}$ is imaginary at every point $\vec{r}$ in space, and the average of the integrals in Eq. (\ref{eq5}) should be considered equal to zero. However, it is possible to show that such reasoning is not rigorous and that part of the integral for $\langle\Delta\vec{v}(t)\rangle$ is never equal to zero even for a single resonant mode due to the presence of a momentum distribution associated with induced currents. To show this, one invokes the field equality (see Appendix Section \ref{eq07derivation} and \cite{boyer2005illustrations})
\begin{equation}\label{eq7}
\int \vec{S}d^3r=\int \vec{r}\vec{J}\cdot\vec{E}d^3r +\int \vec{r}\frac{\partial u(\vec{r},t)}{\partial t} d^3r, 
\end{equation}
derived from Maxwell's equations, where $\vec{J}(\vec{r},t)$ is the current density in $\Gamma$. Substituting Eq. (\ref{eq7}) into Eq. (\ref{eq5}) for $\vec{V}_{CM}=0$ and taking averages one finds
\begin{equation}\label{eq8}
M\langle \Delta\vec{v}\rangle=-{\mathcal{I}^{(J)}_{\Gamma+\Sigma}}-{\mathcal{I}^{(u)}_{\Gamma+\Sigma}}-{\mathcal{I}^{(u)}_{V_0}}
\end{equation}
with 
\begin{equation}\label{eq9new}
{\mathcal{I}^{(J)}_{\Gamma+\Sigma}}= \frac{1}{c^2}\bigg\langle \int_{\Gamma+\Sigma}\vec{r}\vec{J}\cdot\vec{E}d^ 3r \bigg\rangle ,
\end{equation}
\begin{equation}\label{eq11new}
{\mathcal{I}^{(u)}_{V_0}}= \frac{1}{c_0^2}\bigg\langle \int_{V_0}\vec{r} \frac{\partial u (\vec{r},t)}{\partial t} d^3r \bigg\rangle ,
\end{equation}
and
\begin{equation}\label{eq10new}
{\mathcal{I}^{(u)}_{\Gamma+\Sigma}}= \frac{1}{c^2}\bigg\langle \int_{\Gamma+\Sigma}\vec{r} \frac{\partial u (\vec{r},t)}{\partial t} d^3r \bigg\rangle ,
\end{equation}
where the different contributions of each integrand were separated for clarity. 

Eqs. (\ref{eq8})-(\ref{eq10new}) have an apparent physical interpretation when one recognizes each integrand as the product of $\vec{r}$ (with origin at $\vec{R}_{CM}$) and the rate of change of mass density (or ``mass flow'') 
\begin{equation}\label{massflow}
\dot{\rho}_{diss}=\frac{\vec{J}\cdot\vec{E}}{c^2},
\end{equation}
associated with the radiation damping of the induced fields, and
\begin{equation}\label{eq9}
\dot{\rho}_{0EM}=\frac{1}{c_0^2}\frac{\partial u(\vec{r},t)}{\partial t}, \qquad \dot{\rho}_{EM}=\frac{1}{c^2}\frac{\partial u(\vec{r},t)}{\partial t}, 
\end{equation}
as changes in the density of ``free electromagnetic mass'' of the fields in $V_0$ and $\Gamma + \Sigma$, respectively. For non-idealized cavities, the radiation damping term has a non-zero average proportional in principle to the conductivity of the material (that is $\dot{\rho}_{diss}=\sigma {E}^2 /c^2$). Such damping manifests itself as the induction of current shells on the inner side of $\Gamma$ and in the feeder structure of the generator $\Sigma$. 

Moreover, if $\dot{\rho}_{diss}(r)$ has the same distribution as the original mass density $\rho$, the space integral in Eq. (\ref{eq8}) will not contribute to $\Delta\vec{v}$. A sphere, a pill-box or a rectangular cavity are such cases, when storing pure resonant modes resulting in symmetrical dissipative current distributions. Similar considerations of symmetry are valid for the EM density variations in Eqs. (\ref{eq9}). However, they represent energy distributions in the free-space of $V_0$ (Eq. (\ref{eq10new})) or in the conductor (Eq. (\ref{eq11new})). The dependence expressed by the energy  terms in Eq. (\ref{eq9}) indicates that symmetrical distributions do not contribute to the velocity change. 

For an excitation separating the space and time dependencies fully, that is, when one can write $\vec{{E}}(\vec{r},t)=\vec{\mathfrak{E}}(\vec{r})\varepsilon (t)$ and $\vec{{H}}(\vec{r},t)=\vec{\mathfrak{H}}(\vec{r})\eta (t)$, the integral for, e. g., $V_0$ in Eq. (\ref{eq9}) becomes
\begin{equation}\label{eq10}
\bigg\langle\int_{V_0}\vec{r}\frac{\partial u(\vec{r},t)}{\partial t}d^3r \bigg\rangle = \frac{\epsilon_0}{2}\bigg\langle\frac{d}{dt}\varepsilon(t)^2\bigg\rangle\int_{V_0}\vec{r}\mathfrak{E}^2(\vec{r})d^3r+\frac{\mu_0}{2}\bigg\langle\frac{d}{dt}\eta(t)^2\bigg\rangle\int_{V_0}\vec{r}\mathfrak{H}^2(\vec{r})d^3r.
\end{equation}
To assume complete separation is approximately valid in the forced regime by neglecting all cavity modes except a single one with frequency $\omega_0$. In this case, $\varepsilon(t) \propto \sin\omega_0 t$, and $\eta(t) \propto \cos\omega_0 t$. From this we immediately conclude that, if the field configuration is stationary, the contribution from $\dot{\rho}_{EM}$ is zero in $V_0$, $\Gamma$ and $\Sigma$. The remaining non-zero component in Eq. (\ref{eq8}) is the radiation damping term in $\Gamma$ and $\Sigma$
\begin{equation}
M\langle\Delta\vec{v}\rangle=-\frac{\sigma}{2c^2}\bigg\langle\int_{\Sigma+\Gamma}\vec{r}E^2(\vec{r})d^3r \bigg\rangle,
\end{equation}
with
\begin{equation}\label{eq10b}
\bigg\langle \int_{\Sigma+\Gamma}\vec{r}E^2d^3r \bigg \rangle = \bigg\langle\varepsilon(t)^2\bigg\rangle \int_{\Sigma+\Gamma}\vec{r}\mathfrak{E}^2(\vec{r})d^3r,
\end{equation}
for an excitation separating the space and time dependencies. However, before ruling out other possible contributions from the energy densities in $\Gamma$ and $V_0$, it is necessary to assess the driving regimes which could render the time average in Eq. (\ref{eq10}) non-zero as well. 

\section{Analysis of a single mode}

Most references in the theory of microwave cavities \cite{collin1990field,ilinskiui1993propagation} consider the time dependence of the fields as simply as $\varepsilon(t) \propto e^{-i\omega t}$, because they are interested in the space distribution of the modes as a function of the resonator shape and in how these modes are affected by perturbations in the stationary regime. The process of filling a cavity with RF energy is not stationary however. It is most often described using the equivalence between driven RLC circuits and 2-port microwave lossy resonators \cite{karlsson2014microwave, pozar2011microwave}. In the presence of such losses, an approximate time dependent model $\varphi(t)$ for the functions $\varepsilon(t)$ and $\eta(t)$ in Eq. (\ref{eq10}) and a dominant \textit{single resonant mode} with frequency $\omega_0$ obeys the differential relation
\begin{equation}
\ddot{\varphi}+\Gamma_0\dot{\varphi}+ \omega_0^2\varphi=\omega_0^2f_0\cos\omega t ,
\end{equation}
with $\omega$ the generator frequency, $f_0$ a parameter proportional to the generator amplitude, and $\Gamma_0$ the mode-dependent damping coefficient equal to $\omega_0/Q$, with $Q$ the quality factor of the mode. Such equation has the well-known general solution
\begin{equation}\label{eqgen}
\varphi(t)=e^{-\Gamma_0 t/2}\left[\alpha\cos\lambda\omega_0 t+\left(\frac{\Gamma_0\alpha}{2}+\beta\right)\frac{\sin\lambda\omega_0t}{\lambda\omega_0}\right]+f_0\left[\frac{(\omega_0^2-\omega^2)\cos\omega t+\Gamma_0\omega\sin\omega t}{D}\right],
\end{equation}
where
\begin{equation}
\begin{split}
\alpha &=\varphi(0)-f_0(\omega_0^2-\omega^2)/D, \quad
\beta =\dot{\varphi}(t)-\frac{f_0\Gamma_0\omega^2}{D},\\
\lambda &=\sqrt{1-\left(\Gamma_0/2\omega_0\right)^2},\quad
D =\frac{(\omega_0^2-\omega^2)^2+\Gamma_0^2\omega^2}{\omega_0^2},
\end{split}
\end{equation}
and $\varphi(0)$ and $\dot{\varphi}(0)$ represent the field initial state. At resonance, $\omega=\omega_0$, and for $\varphi(0)=0$ and $\dot{\varphi}(0)=0$, the general solution becomes
\begin{equation}\label{eqress}
\varphi_{\omega\rightarrow\omega_0}(t)=f_0\frac{\omega_0}{\Gamma_0}\left[\sin\omega_0 t-\frac{e^{-\Gamma_0 t/2}}{\lambda}\sin\lambda\omega_0t\right].
\end{equation}
The limit of an ideal cavity is obtained by  setting $\Gamma_0\rightarrow 0$ (or $Q\rightarrow\infty$). To first order in $\Gamma$, an approximation for real cavities results in
 \begin{equation}\label{eqress2}
\varphi_{\omega\rightarrow\omega_0}(t) \approx \omega_0 f_0\left\{\frac{1}{2}t\sin\omega_0t-\frac{\Gamma_0}{8}\left[\left(\frac{1}{\omega_0^2}+t^2\right)\sin\omega_0t-\left(\frac{t}{\omega_0}\right)\cos\omega_0t\right]\right\}.
\end{equation}
Therefore, disregarding the $\Gamma_0$-dependence at resonance, the electric field amplitude increases, for example, as $\varepsilon(t)=(f_0/2)\omega_0t\sin\omega_0 t$, and its average value is not zero during the transient regime.

\subsection{Time averages} \label{sec3.2}

Because the fields are oscillatory, the configurations rendering the result of Eq. (\ref{eq5}) different from zero require that the average value of quantities proportional to $\vec{\mathfrak{E}}\times\vec{\mathfrak{H}}$  be non-zero as well. It is sufficient to calculate averages as given by
\begin{equation}\label{eq6b}
\langle\Delta\vec{v}(t)\rangle=\int_{-\infty}^{\infty}w(t,t')\Delta \vec{v}(t')dt',
\end{equation}
with $w(t,t')$ a weight function, and $\pm\infty$ a representative limit of measurement time intervals much longer than $\omega^{-1}$. The average is justified because in the measurement process of such velocity change (involving, in general, a mechanical set-up to measure torsions, see, e. g. \cite{white2017, tajmarspace}), the output is mechanically unresponsive to oscillations at the microwave frequencies. Such time average is accomplish by considering, for example, the weighting function 
\begin{equation}
w(t,t')=\frac{1}{2\pi}\int_{-\infty}^{\infty}\Xi(\omega,\langle\omega\rangle)e^{i\omega(t-t')}d\omega,
\end{equation}
with $\Xi(\omega,\langle\omega\rangle)$ a Gaussian function centered at $\langle\omega\rangle$. Then, the average is calculated for $\langle\omega\rangle=0$, which corresponds to extracting the DC component of the function $\Delta \vec{v}(t)$.

In line with such reasoning, one uses the weighting function
\begin{equation}\label{smoothfunc}
w(t,t')=\frac{1}{\tau\sqrt{2\pi}}e^{-(t'-t)^2/2\tau^2},
\end{equation}
\begin{figure}[ht]
\centering
  \includegraphics[width=160mm]{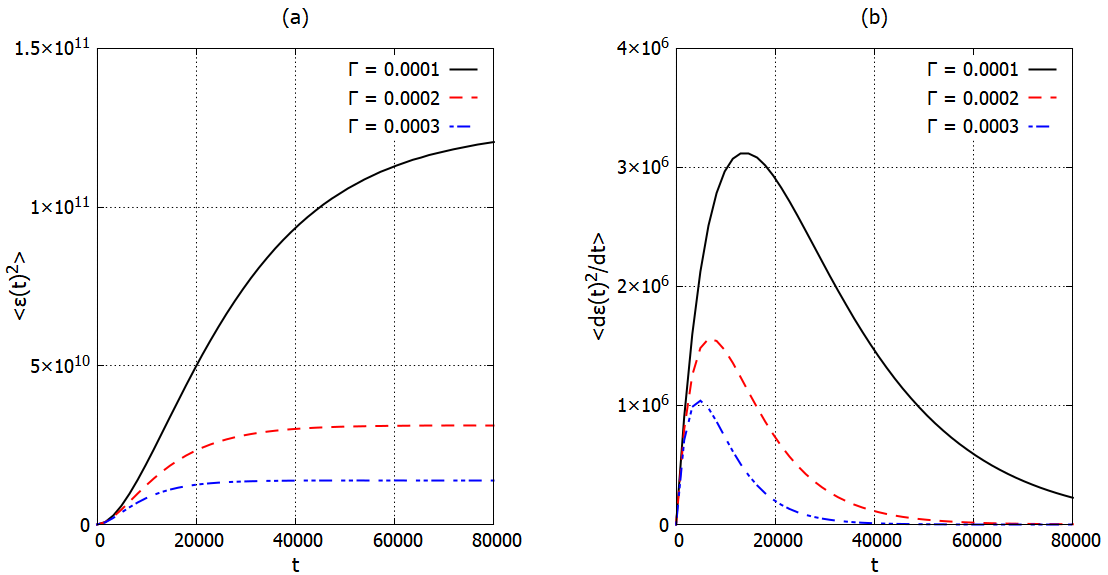}
  \caption{Time dependence of the average values for the principal components of the momentum integrals, Eq. (\ref{eq24_0}) and (\ref{eq24}), in arbitrary units of time for the specified values of $\Gamma_0$.}
  \label{fig:02} 
\end{figure}
so that the time averages in Eq. (\ref{eq10}) and (\ref{eq10b}) for the resonance expresions, Eq. (\ref{eqress}) when $\omega_0\tau \gg 1$, are given by 
\begin{equation}\label{eq19}
\begin{split}
\bigg\langle\varepsilon(t)^2\bigg\rangle &=\frac{f_0^2\omega_0^2}{\Gamma_0^2}\left\{ 1+e^{-\Gamma_0 t}-2\left[1-\frac{1}{2}\left(\frac{\Gamma_0^2t}{8\omega_0}\right)^2\right]e^{-\Gamma_0 t/2}\right\},\\
\bigg\langle\frac{d}{dt}\varepsilon(t)^2\bigg\rangle &=\frac{f_0^2\omega_0^2}{2\Gamma_0}\left[1-\left(\frac{\Gamma_0^2t}{8\omega_0}\right)^2-e^{-\Gamma_0 t/2}\right]e^{-\Gamma_0 t/2},
\end{split}
\end{equation}
(see Appendix \ref{apresfunc}), with similar expressions for the magnetic counterparts. The result of Eqs. (\ref{eq19}) indicates that one can express them as a function of the mode quality factor
\begin{equation}\label{eq25b}
\begin{split}
\bigg\langle \epsilon(t)^2\bigg\rangle &=Q^2f_0^2 a_E(t),\\
\bigg\langle \frac{d}{dt}\epsilon(t)^2\bigg\rangle &=\omega_0 Qf_0^2 b_E(t),
\end{split}
\end{equation}
with $a_E(t)$ and $b_E(t)$ nondimensional numbers as those within the brackets of Eqs. (\ref{eq19}).

Figure \ref{fig:02} shows plots of the average integrals Eqs. (\ref{eq24_0}) and (\ref{eq24}) for different values of $\Gamma_0$. The function given by Eq. (\ref{eq24}) has a maximum at  $t_{max}\approx 2\ln2/\Gamma_0$. At this time
\begin{equation}\label{eq21}
\bigg\langle\frac{d}{dt}\varepsilon(t_{max})^2\bigg\rangle =\frac{f_0^2\omega_0^2}{4\Gamma_0}\left[1-\left(\frac{\Gamma_0\ln 2}{4\omega_0}\right)^2\right].
\end{equation}
Physically meaningful values of $t_{max}$ control the rate of momentum change and decide the onset of any observable reaction of the mechanical subsystem as described in Section \ref{momentumvariation}.

\subsection{Estimating the stored EM momentum}\label{estimatingmomentum}

Assuming excitation of a single mode, the contribution from $\vec{J}$ in Eq. (\ref{eq9new}) requires the determination of the electric field $\vec{E}_{\Gamma}$ in the region $\Gamma$. Following \cite{stratton2007electromagnetic}, one uses the reference system $(\hat{\xi},\hat{\eta},\hat{\zeta})$ with $\hat{\zeta}$ an unitary outward vector orthogonal to the inner cavity surface. In terms of the magnetic field $\vec{H}_{||}$ parallel to the surface in $V_0$, the current  density in the conductor is approximated as
\begin{equation}\label{eqcurdensity}
\vec{J}=\left(\frac{\omega}{c}\right)\frac{\gamma}{\sqrt{1+i\gamma}}\vec{H}_{||}e^{-ik\vec{\zeta}\cdot\delta \vec{r}},
\end{equation} 
with $\gamma =\sigma/\epsilon\mu$, $\vec{\delta r} = \vec{r}-\vec{r}_{\Gamma}$ the distance of a point in $\Gamma$ from the surface at $\vec{r}_{\Gamma}$, and $\vec{k}=k\hat{\zeta}$ the wave vector within the metal. Therefore, the ``mass-flow'' term, Eq. (\ref{massflow}), results in 
\begin{equation}
\dot{\rho}_{diss}=\left(\frac{\omega}{c^2}\right)\frac{\gamma}{\sqrt{1+\gamma^2}}\mu H^2_{||}e^{2\Im (-ik\vec{\zeta}\cdot\delta\vec{r})},
\end{equation}
with $\Im (x)$ the imaginary part of $x$. 
\begin{figure}[ht]
\centering
 \includegraphics[width=0.35\columnwidth]{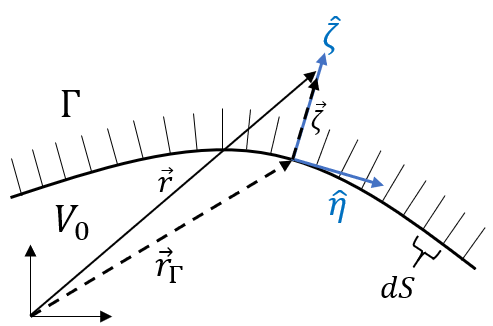}
  \caption{An evanescent field exists in region $\Gamma$, propagating along $\hat{\zeta}$ at point $\vec{r}_{\Gamma}$. The integral of the momentum density is calculated as an integral along $\hat{\zeta}$ (orthogonal to the surface) and another surface integral for all inner cavity area elements $dS$ at $\vec{r}_{\Gamma}$.}
  \label{fig:05} 
\end{figure}

Given the definition of $\vec{k}$ in accordance with \cite{stratton2007electromagnetic}, the idea is to obtain an approximate expression for Eq. (\ref{eq9new}) by taking the integral along $\hat{\zeta}$ first (orthogonal to the surface as depicted in Figure \ref{fig:05}), and then summing all contributions from the element of area $dS$. This idea is justified if the minimum curvature radius of the inner surface is larger than the typical skin depth in the conductor. This is the case for cooper, for which $\gamma = 5.4\times 10^8$ at 2 GHz and the skin depth is about 2.0 $\mu$m as cited previously. Since the area of the cavity feeder can be considered much smaller than the total surface $\Gamma$, one neglects the integration over $\Sigma$. Writing the position vector in $\Gamma$ as $\vec{r}=\vec{r}_{\Gamma}+\zeta\hat{\zeta}(\vec{r}_{\Gamma})$, one can then approximate
\begin{equation}\label{eq39b}
\frac{1}{c^2}\int_{\Gamma+\Sigma}\vec{r}(\vec{J}\cdot\vec{E})d^3r\approx\left(\frac{\mu}{c}\right)\left[\frac{1}{\sqrt{2\gamma}}\oint_{\Gamma}\vec{r}_{\Gamma}H^2_{||}(\vec{r}_{\Gamma})dS+{\frac{\lambda_0}{\gamma}}\oint_{\Gamma}H^2_{||}(\vec{r}_{\Gamma})\hat{dS}\right].
\end{equation}
In Eq. (\ref{eq39b}), $\lambda_0$ is the EM wavelength in free space, and $\hat{dS}=\hat{\zeta}(\vec{r}_{\Gamma})dS$. Using the single mode separation and the previous result in Eqs. (\ref{eq19}) and (\ref{eq25b}) for $\eta(t)$, Eq. (\ref{eq9new}) can be expressed as
\begin{equation}\label{eq41}
{\mathcal{I}^{(J)}_{\Gamma+\Sigma}} \approx Q^2\left(\frac{\mu g_0^2}{c_0}\right)a_H(t)\left[\frac{1}{\sqrt{2\gamma}}\oint_{\Gamma}\vec{r}_{\Gamma}{\mathfrak H}^2_{||}(\vec{r}_{\Gamma})dS+{\frac{\lambda_0}{\gamma}}\oint_{\Gamma}{\mathfrak H}^2_{||}(\vec{r}_{\Gamma})\hat{dS} \right],
\end{equation}
with $g_0$ the amplitude of the input magnetic field. Eq. (\ref{eq41}) shows that the momentum contribution of the surface current is proportional to $Q^2$ and is the product of the input momentum density $\mu g_0^2/c$ by a vector integral (given by an integral of the parallel magnetic field over the cavity surface) with unit of volume. The second integral (proportional to $\lambda_0$) is much smaller than the main term proportional to $\vec{r}_{\Gamma}$ and will be neglected.

Using Eqs. (\ref{eq25b}) and the separation by Eq. (\ref{eq10}), the momentum field contribution in $V_0$, Eq. (\ref{eq11new}), is written for clarity as
\begin{equation}\label{eq42}
{\mathcal{I}^{(u)}_{V_0}}=\frac{Q}{4\pi}\left[b_E(t)\left(\frac{\epsilon_0 f_0^2}{c_0}\right)\left(\frac{1}{\lambda_0}\int_{V_0}\vec{r}{\mathfrak E}^2(r)d^3r\right) +b_H(t)\left(\frac{\mu_0 g_0^2}{c_0}\right)\left(\frac{1}{\lambda_0}\int_{V_0}\vec{r}{\mathfrak H}^2(r)d^3r\right)\right].
\end{equation}
The momentum contribution of the fields in $V_0$ is proportional to $Q$ and, for each field component, is the product of the momentum density ($\epsilon_0f_0^2/c_0$ and $\mu_0 g_0^2/c_0$ for the electric and magnetic fields, respectively) and a vector integral with unit of volume. 

Finally, the last contribution of the EM energy in the cavity body, or Eq. (\ref{eq10new}), is also easily estimated using the same approximation for the electric field in the conductor as due to the an incident parallel magnetic contribution arising from for the current  density Eq. (\ref{eqcurdensity}). The solution is given entirely as a function of the magnetic energy
\begin{equation}\label{eq32b}
{\mathcal{I}^{(u)}_{\Gamma+\Sigma}}\approx\frac{Q}{2\pi^2}b_H(t)\left(\frac{\mu g_0^2}{2c_0}\right)\left[\sqrt{\frac{2}{\gamma}}\oint_{\Gamma}\vec{r}_{\Gamma}{\mathfrak H}_{||}^2dS+\left({\frac{2\lambda_0}{\pi\gamma}}\right)\oint_{\Gamma}{\mathfrak H}_{||}^2\hat{dS} \right].
\end{equation}
Again, the contribution to the EM momentum of the energy density in $\Gamma$ is the sum of two terms proportional to Q and the momentum density similarly to Eq. (\ref{eq41}). Note that the outcome of Eq. (\ref{eq32b}) shares the same origin of Eq. (\ref{eq42}), however, it is the result of an integration over a much smaller volume (reduced by the factor $\gamma^{-1/2}$), whose size is controlled by the microwave skin depth. Since $\gamma^{-1/2}=4.3\times 10^{-5}$, one may neglect the second terms (depending on $\hat{\zeta}(\vec{r}_{\Gamma})$ and $\gamma^{-1}$) for the sake of estimating the main contributions of the surface integrals in Eqs. (\ref{eq41}) and (\ref{eq32b}). 

\subsection{Space averages: cylinder and conical frustum illustration}\label{condspace}

\begin{figure}[ht]
\centering
 \includegraphics[width=60mm]{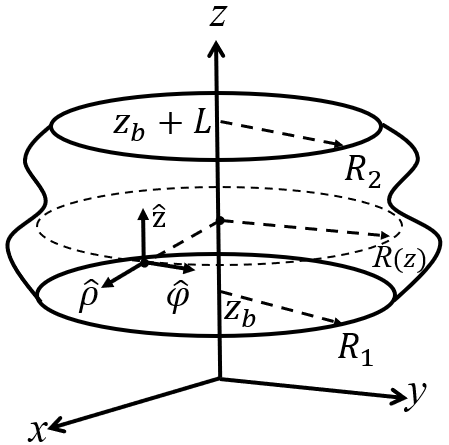}
  \caption{A general BOR cavity defined by the rotation of the function $R(z)$ terminated by two flat covers of radii $R_1$ and $R_2$ located at $z_b$ and $z_b+L$, respectively.}
  \label{fig:04a} 
\end{figure}

In the light of Eqs. (\ref{eq41}), (\ref{eq42}) and (\ref{eq32b}), one rewrites the integrals of the space averages $\langle \vec{r}{\mathfrak E}^2\rangle$, $\langle \vec{r}{\mathfrak H}^2\rangle$ and $\langle \vec{r}{\mathfrak H_{||}}^2\rangle_{\Gamma}$ as a function of cavity geometry and the position of the initial CE. If one considers body-of-revolution (BOR) cavities, the original center will be located along the symmetry axis. The conical frustum is such a case, but other shapes are possible. This is illustrated in Figure \ref{fig:04a}, depicting an arbitrary truncated BOR cavity terminated in two flat covers . The space averages in Eqs. (\ref{eq41}), (\ref{eq42}) and (\ref{eq32b}) are redefined explicitly in terms of the energy averages:
\begin{equation}\label{eq22b}
\begin{split}
\langle \vec{r}{\mathfrak E}^2\rangle &= (\vec{R}_{E}-\vec{R}^{(V)}_{CM})\int_{V_0} {\mathfrak E}^2(\vec{r})d^3r,\\
\langle \vec{r}{\mathfrak H}^2\rangle &= (\vec{R}_{H}-\vec{R}^{(V)}_{CM})\int_{V_0} {\mathfrak H}^2(\vec{r})d^3r,\\
\langle \vec{r}{\mathfrak H_{||}}^2\rangle_{\Gamma} &=(\vec{R}_{H_\Gamma}-\vec{R}^{(S)}_{CM})\oint_{\Gamma} {\mathfrak H}_{||}^2(\vec{r}_{\Gamma})dS,
\end{split}
\end{equation}
with 
\begin{equation}\label{eqs23}
\begin{split}
\vec{R}_{E}\int {\mathfrak E}^2(\vec{r})d^3r &=\int_{V_0} \vec{r}{\mathfrak E}^2(\vec{r})d^3r,\\
\vec{R}_{H}\int {\mathfrak H}^2(\vec{r})d^3r &=\int_{V_0} \vec{r}{\mathfrak H}^2(\vec{r})d^3r,\\
\vec{R}_{H_\Gamma}\oint_{\Gamma} {\mathfrak H}_{||}^2(\vec{r_{\Gamma}})dS &=\oint_{\Gamma} \vec{r}_{\Gamma}{\mathfrak H}_{||}^2(\vec{r}_{\Gamma})dS
\end{split}
\end{equation}
defining the positions of $\vec{R}_E$, $\vec{R}_H$ as the energy centroids for the electric and magnetic fields in $V_0$, respectively, and $\vec{R}_{H_{\Gamma}}$ the centroid of the magnetic contribution in $\Gamma$. For a BOR cavity of length $L$, whose sides are described by a function $R(z)$ as shown in Figure \ref{fig:04a}, with $z\in\{0,z_L\}$ ($z_b=0$), the CE is obtained from Eqs. (\ref{eqs23}) assuming constant ${\mathfrak E}^2$ or ${\mathfrak H}^2$, as located initially at 
\begin{equation}\label{eq25}
\vec{R}^{(V)}_{CM}=\hat{z}\frac{\int_{0}^{L} z R(z)^2 dz}{\int_{0}^{L} R(z)^2 dz},\; \;\;\vec{R}^{(S)}_{CM}=\hat{z}\frac{LR_2^2+2\int_{0}^{L} [z R(z)/\hat{\zeta}(z)\cdot\hat{\rho}] dz}{(R_1^2+R_2^2)+2\int_{0}^{L}[ R(z)/\hat{\zeta}(z)\cdot\hat{\rho}] dz}, 
\end{equation}
where $\hat{\zeta}(z)$ is the unitary vector normal to the surface at $z$ (Figure \ref{fig:05}).

For a vacuum filled cylinder with radius $R$ and length $L$ ($R(z)=R, \;\; \hat{\zeta}(z)\cdot\hat{\rho}=1$), one approximates the fields with the values obtained analytically from an ideal model with \textit{perfectly conducting} walls. The TM fields, for instance, in a coordinate system with center at cylinder geometric centroid ($\vec{R}^{(V)}_{CM}=\vec{R}^{(S)}_{CM}=\hat{z}L/2$), are written as \cite{zangwill2013modern}
\begin{equation}\label{eq22}
\begin{split}
\vec{E}(\vec{r},t)&=2\varepsilon(t)\left[\frac{n\pi}{L\gamma_{\omega}^2}\vec{\nabla}_{\perp}\psi\sin \frac{p\pi z}{L}+\hat{z}\psi\cos \frac{p\pi z}{L}\right],\\
\vec{H}(\vec{r},t)&=\eta(t)\frac{i\omega\epsilon}{\gamma_{\omega}^2}\left( \hat{z}\times\vec{\nabla}_{\perp}\psi \right)\cos \frac{p\pi z}{L},
\end{split}
\end{equation}
with $z\in\{0,L\}$, $p$ a non-negative integer number, $\gamma_{\omega} =\sqrt{\mu\epsilon\omega^2-p^2\pi^2/L^2}$, $\vec{\nabla}_{\perp}=\hat{\rho}\partial/\partial\rho+\hat{\phi}\partial/\partial\phi$ and $\psi$ is a solution of the equation
\begin{equation}\label{eq37b}
\left[\frac{1}{\rho}\frac{\partial}{\partial\rho}\left(\rho\frac{\partial\psi}{\partial\rho}\right)+\frac{1}{\rho^2}\frac{\partial^2\psi}{\partial\varphi^2}+\gamma_{\omega}^2\psi\right]=0,
\end{equation}
subjected to the boundary condition $\psi(R,\varphi)=0$. Such solutions are of the form $\psi(\rho,z)=J_m(\rho)e^{\pm im\varphi}$ where $J_m(\rho)$ are the Bessel functions \cite{jackson1999classical} of order $m$, with $m=\{0,1,2,\ldots\}$ .

Using the cylinder center as the origin of the reference system, the magnetic averages in Eqs. (\ref{eq22b}) become
\begin{equation}\label{eqs36b}
\begin{split}
\langle \vec{r}{\mathfrak H}^2\rangle &\propto \hat{z}\int_{0}^{R}\left(\frac{\partial J_m}{\partial\rho}\right)^2\rho d\rho\int_{-L/2}^{L/2}z\cos^2 p\pi\left(\frac{z}{L}+\frac{1}{2}\right)dz=0,\\
\langle \vec{r}_{\Gamma}{\mathfrak H}_{||}^2\rangle &\propto 2\pi\hat{z}\left(\frac{\partial J_m}{\partial \rho}\right)^2\bigg\rvert_{\rho=R}\int_{-L/2}^{L/2}z\cos^2 p\pi\left(\frac{z}{L}+\frac{1}{2}\right)dz=0,
\end{split}
\end{equation}
because the distribution of magnetic energy is $\varphi$-independent and symmetric about the center. In writing the second equation in Eqs. (\ref{eqs36b}), the integrals over the cylinder covers
\begin{equation}\nonumber
\pm\hat{z}\pi L\int_{0}^{R}\left(\frac{\partial J_m}{\partial \rho}\right)^2\rho d\rho
\end{equation}
cancel out, because they point in opposite directions. By using the first equation in Eq. (\ref{eq22}), it is easily seen that
\begin{equation}
\langle \vec{r}{\mathfrak E}^2\rangle =0,
\end{equation}
with a similar result for all TE modes. For this reason, there is no contribution of the EM momentum in the velocity change in Eq. (\ref{eq8}), a conclusion that does not depend on the assumption of a single mode, but on the cylindrical symmetry of all modes. 

For the conical frustum, one defines
\begin{equation}
\xi=(R_2-R_1)/R_1, \quad u=z/L, \quad \rho_1=R_1/L, \quad R(u)=L\rho_1(1+u\xi),
\end{equation}
where $R_1, R_2$  are the frustum radii at $z_b$ and $z_b+L$, respectively. The integrals in Eq. (\ref{eq25}) with $z_b=0$ result in the distinct centers
\begin{equation}\label{eq28}
\vec{R}^{(V)}_{CM}=\hat{z}L \left[
\frac{1/2+2\xi/3+\xi^2/4}{1+\xi+\xi^2/3}\right], \;\;\;\vec{R}^{(S)}_{CM}=\hat{z}L\Bigg\{\frac{\rho_1(1+\xi)^2+(2\xi/3+1)\sqrt{\xi^2\rho_1^2+1}}{\rho_1[1+(1+\xi)^2]+(\xi+2)\sqrt{\xi^2\rho_1^2+1}}\Bigg\}.
\end{equation}
If $R_1=R_2$, the above equations return $\vec{R}^{(V)}_{CM}=\vec{R}^{(S)}_{CM}=\hat{z}L/2$, or the center of the cylinder treated previously. 

\begin{figure}[t!]
\centering
 \includegraphics[width=0.9\columnwidth]{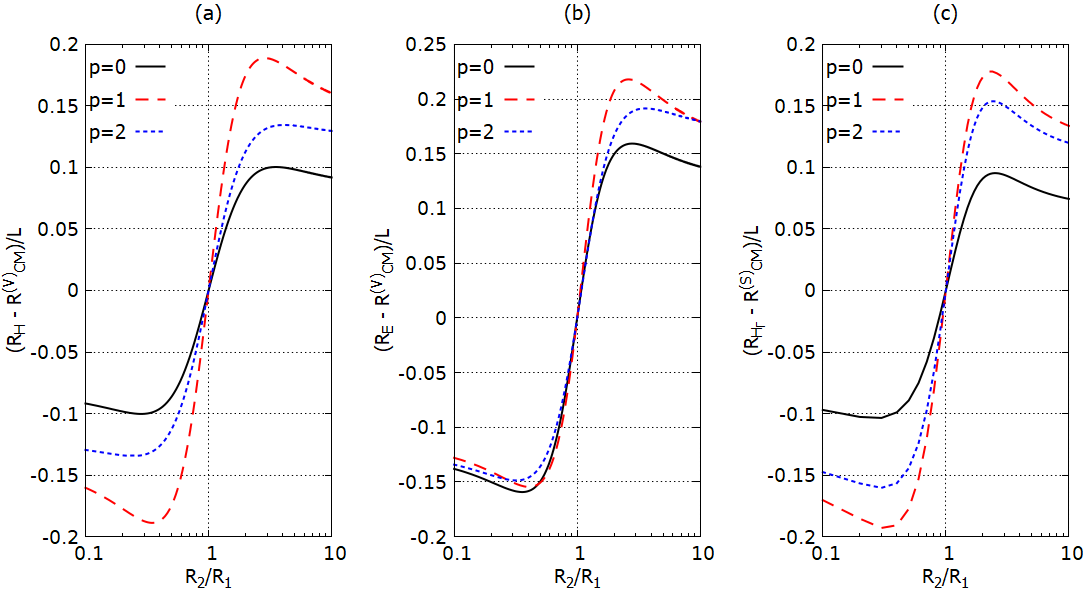}
  \caption{Asymmetry in the position of the CE, as given by Eqs. (\ref{eq40}). As a function of the frustum cover ratio $R_2/R_1$ for the fields generated by the illustration Eq. (\ref{eq31}): contribution from the magnetic (a), electric (b) and surface magnetic (c) volume integrals.}
  \label{fig:04} 
\end{figure}

When $R_1\neq R_2$, it is not possible to solve Eq. (\ref{eq37b}) analytically, and no closed relation exists for the modes in Eqs. (\ref{eq22}). However, nothing prevent us of assuming an analytical function $\psi(\rho,z)$  in order to illustrate the outcome of the space average for an asymmetrical energy distribution. Therefore, as an \textit{illustration} one may set
\begin{equation}\label{eq31}
\psi(u,s)=\frac{\psi_0}{2}[\rho_1^2(1+u\xi)^2-s^2],
\end{equation}
where $s=\rho L$ and $\psi_0$ a constant. Such relation is a polynomial approximation for $J_0(\bar{x})=0$ when $\bar{x}\approx 2.405$ for all points on the side $s(u)=\rho_1(1+u\xi)$. Eqs. (\ref{eq22b}) for the magnetic field become
\begin{equation}
\begin{split}
\langle \vec{r}{\mathfrak H}^2\rangle &=\hat{z} \frac{\pi}{2} L^4H_0^2\rho_1^4\left[\mathcal{I}_{C2}(p,\xi)^{(4,2)}-\mathcal{I}_{C1}(p,\xi)^{(4,2)}R^{(V)}_{CM}/L\right],\\
\langle \vec{r}{\mathfrak H}_{||}^2\rangle &=\hat{z} \frac{\pi}{2} L^4H_0^2\rho_1^4\Bigg\{ \rho_1\left[(1+\xi)^4(1-R_{CM}^{(S)}/L)-R_{CM}^{(S)}/L\right]\Bigg. \\
&+\Bigg. 4\sqrt{\xi^2\rho_1^2+1}\left[I_{C1}(p,\xi)^{(3,2)}-I_{C2}(p,\xi)^{(3,2)}R_{CM}^{(S)}/L\right]\Bigg\},
\end{split}
\end{equation}
with the integrals $\mathcal{I}_{C1}(p,\xi)^{(4,2)}$, $\mathcal{I}_{C2}(p,\xi)^{(4,2)}$, $I_{C1}(p,\xi)^{(3,2)}$, and  $I_{C2}(p,\xi)^{(3,2)}$ defined in Eqs. (\ref{eq39}) of Appendix \ref{centerenergy}, $H_0=2\omega\epsilon_0 \psi_0/(\gamma^2 L)$, $R_{CM}^{(V)}$ and $R_{CM}^{(S)}$ given by Eq. (\ref{eq28}). The result depends on the integer $p$ giving the number of zeros along $z$ in the energy distribution. The effect of the space average is best seen by calculating the new position of $\vec{R}_H$ and  $\vec{R}_E$ as defined in Eqs. (\ref{eqs23}). For $p=0$ ($z_b=0$), the assumption facilitates the calculation of the new center exactly, and it is possible to show that
\begin{equation}
\begin{split}
\vec{R}_{H}&=\vec{R}_{E}=\hat{z}L\left[\frac{1/2+4\xi/3+3\xi^2/2+4\xi^3/5+\xi^6/6}{1+2\xi+2\xi^2+\xi^3+\xi^4/5}\right],\\
\vec{R}_{H_{\Gamma}} &=\hat{z}L\Bigg\{\frac{\rho_1(1+\xi)^4)+4\sqrt{\xi^2\rho_1^2+1}(1+4\xi+3\xi^2+4\xi^3/5)}{\rho_1[1+(1+\xi)^4]+\sqrt{\xi^2\rho_1^2+1}(4+6\xi+4\xi^2+\xi^3)}\Bigg\}.
\end{split}
\end{equation}

The position of the displaced centers, as a function of $1+\xi=R_2/R_1$, is seen in Figure \ref{fig:04} for some values of $p$, $\rho_1=0.611$ and the ratio between the transversal and longitudinal field components equal to 0.2. The limits $R_2/R_1\rightarrow\infty$ and $R_2/R_1\rightarrow 0$ correspond to the cone. The displacement exhibit a maximum of $\approx L/5$ (e. g., for the ratio $ R_2/R_1 \approx 2.5$, and $p=1$). 
 
\newpage
\subsection{Momentum exchange in the transient regime}
\label{momentumvariation}

Lorentz forces provide the interface for the momentum exchange between the fields and the cavity body. The resultant of such forces are equivalent to a thrust of internal origin acting on the mechanical subsystem to satisfy momentum conservation. Such thrust is decomposable as factors of Eq. (\ref{eq8}):
\begin{equation}
\vec{{\mathfrak F}}_{\Gamma+\Sigma}^{(J)}=-\frac{\partial {\mathcal{I}^{(J)}_{\Gamma+\Sigma}}}{\partial t},\;\;\;\;\;\;\vec{{\mathfrak F}}_{V_0}^{(u)}=-\frac{\partial {\mathcal{I}^{(u)}_{V_0}}}{\partial t},\;\;\;\;\;\;\vec{{\mathfrak F}}_{\Gamma+\Sigma}^{(u)}=-\frac{\partial {\mathcal{I}^{(u)}_{\Gamma+\Sigma}}}{\partial t}.
\end{equation}
Given Eqs. (\ref{eq41}), (\ref{eq42}), (\ref{eq32b}) and definition (\ref{eq22b}), such factors can be rewritten as
\begin{equation}\label{eq46}
\begin{split}
\vec{{\mathfrak F}}_{\Gamma+\Sigma}^{(J)} &=-\frac{Q^2}{c_0\sqrt{2\gamma}}\left(\frac{\vec{R}_{H_\Gamma}-\vec{R}^{(S)}_{CM}}{\lambda_0}\right)\Bigg\{\dot{a}_H\mu g_0^2\lambda_0\oint_{\Gamma} {\mathfrak H}_{||}^2(\vec{r}_{\Gamma})dS\Bigg\},\\
\vec{{\mathfrak F}}_{V_0}^{(u)} &=-\frac{Q}{4\pi c_0}\left(\frac{\vec{R}_{E}-\vec{R}^{(V)}_{CM}}{\lambda_0}\right)\Bigg\{\dot{b}_E\epsilon_0 f_0^2\int_{V_0} {\mathfrak E}^2(\vec{r})d^3r\Bigg\}\\ &\;\;\;-\frac{Q}{4\pi c_0}\left(\frac{\vec{R}_{H}-\vec{R}^{(V)}_{CM}}{\lambda_0}\right)\Bigg\{\dot{b}_H\mu_0 g_0^2\int_{V_0} {\mathfrak H}^2(\vec{r})d^3r\Bigg\},\\
\vec{{\mathfrak F}}_{\Gamma+\Sigma}^{(u)} &=-\frac{Q}{2\pi^2 c_0\sqrt{2\gamma}}\left(\frac{\vec{R}_{H_\Gamma}-\vec{R}^{(S)}_{CM}}{\lambda_0}\right)\Bigg\{\dot{b}_H\mu g_0^2\lambda_0\oint_{\Gamma} {\mathfrak H}_{||}^2(\vec{r}_{\Gamma})dS\Bigg\}.
\end{split}
\end{equation}
These forces point towards the shape factors $\vec{\varrho}_{field}=-(\vec{R}_{field}-\vec{R}_{CM})/\lambda_0$ depending on the final energy distribution of the particular excited mode. For the illustrations of Figure (\ref{fig:04}), when $R_2<R_1$, the initial thrust points toward $+\hat{z}$.

Moreover, for all terms in Eqs. (\ref{eq46}), the expressions in the curly brackets are proportional to the source input power, $P_0$. Consequently, the amplitude of the thrust-to-power ratio ${\mathfrak F}/P_0$ for the current damping term follows the proportionality relation  
\begin{equation}\label{eq47b}
{\mathfrak F}_{\Gamma+\Sigma}^{(J)} /P_0 \propto \frac{Q^2}{c_0\sqrt{2\gamma}}|\varrho_{H_\Gamma}|.
\end{equation}
The ratio is therefore modified by the amount $Q^2/\sqrt{2\gamma}$ times the shape factor.  Similarly, according to Eqs. (\ref{eq46}), for $\vec{{\mathfrak F}}_{V_0}^{(u)}$ and $\vec{{\mathfrak F}}_{\Gamma+\Sigma}^{(u)}$, the factor is $Q/4\pi$ and $Q/2\pi^2\sqrt{\gamma}$, respectively. Because $c_0^{-1}=3.33\;\mu$N/kW, and assuming $Q=$ 20000, $\gamma(@1.5 \textrm{GHz})=7.14\times 10^8$, one finds 
\begin{equation}\label{eq48}
\begin{split}
{\mathfrak F}_{\Gamma+\Sigma}^{(J)} /P_0 &\propto 35.3\;|\varrho_{H_\Gamma}|\;\;\textrm{mN/kW}, \\
{\mathfrak F}_{V_0}^{(u)} /P_0 &\propto 5.3\;|\varrho_{E}|\;\;\textrm{mN/kW},\\
{\mathfrak F}_{\Gamma+\Sigma}^{(u)} /P_0 &\propto 0.14\;|\varrho_{H_\Gamma}|\;\;\mu\textrm{N/kW},
\end{split}
\end{equation}
which gives approximate values for the relative intensity of each contribution in Eqs. (\ref{eq46}) for the momentum variation. Based on the calculated amplitude of $\varrho_{field}$ in Section \ref{condspace}, one expects the thrust-to-power ratio to be no larger than one order of magnitude less than the values in Eqs. (\ref{eq48}).

Most importantly is the expected \textit{time range} of such momentum variation. This time is of the order $t_{max}$ calculated in Section \ref{sec3.2} or
\begin{equation}\label{eqlast}
t_{max}=2\ln 2\left(\frac{Q}{\omega}\right).
\end{equation}
Using the same quality factor, for $1.5$GHz, one finds $t_{max}=2.94\;\mu$s. Such time is incompatible with the several seconds observed in many experimental tests \cite{white2017, tajmarspace, yang2016thrust} claiming to proof or disproof the effect. The later situation is probably the case, because after $t_{max}$ no other thrust is expected to occur in the system which nevertheless is under the action of other perturbation forces neglected here. One notes $t_{max}$ will increase with Q as well as the magnitude of the internal thrust as determined by Eqs. (\ref{eq46}).

\section{Conclusions}\label{conclusions}

This work shows that, despite the oscillating fields, there is a non-vanishing contribution of electromagnetic momentum density within a closed microwave cavity carrying a coherent EM field. The main component of this density should be considered when calculating the total momentum of the system. It is chiefly due to strong induced currents in the cavity body created by the penetrating fields  and revealing the significant role played by dissipation in the production of the effect. Yet, other EM momentum averages exist during the transient phase of field growth or decline and contribute to momentum variation as well. In the same way, Pugh and Pugh \cite{pugh1967physical} proposed in 1967 that a static charged sphere enclosing a strong magnetostatic field can be set in motion by the conversion of stored EM momentum after the fields are turned off. Another example is Graham and Laholz experiment \cite{graham1980observation}, where the authors further concluded that ``vacuum is the seat of something in motion whenever static fields are set up with non-vanishing Poynting vector''. In the present case, a net Poynting vector is generated by the oscillating fields inside the metal of the cavity body which acts as a DC momentum.

Besides non-zero time averages, in order to exist such net EM momentum demands non-zero space averages of the distribution of EM energy about the initial CE. These distributions are of two types: (\textit{i}) an integral over the surface magnetic energy as given by Eq. (\ref{eq9new}) and reduced to Eq. (\ref{eq41}); and (\textit{ii}) two integrals, Eq. (\ref{eq11new}) and (\ref{eq10new}), reduced to the Eq. (\ref{eq42}) and (\ref{eq32b}) over the volume energy content in the $V_0$ and $\Gamma$, respectively. Space averages are created by an unbalance flow of the Poynting vector when the cavity shape is not symmetrical in relation to the original CM. Therefore, the intensity of the final stored EM momentum will depend on the excited resonant mode through the quality factor $Q$ and shape factors given by Eqs. (33). 

The presence of a net EM momentum should manifest itself as a change in the mechanical momentum of the cavity body. Following \cite{penfield1966electrodynamics}, one may say that the mechanical subsystem is not closed in relation to its EM counterpart so that any change in the EM momentum would cause a displacement of the mechanical body. But the dynamics is severely restricted because the modified system CE remains contained in $V_0$ and  very close to the original CM. 

The determination of the total stored EM momentum facilitates the calculation of the internal total force as the interface between the EM and the mechanical subsystems. The action of this force (which is not ``instantaneous'') is simply the time derivative of the EM momentum integral at every time $t\geq t_0$. It is significant to emphasize that, at all times, the total momentum as given by Eq. (\ref{eq4}) is rigorously conserved for an isolated and free cavity. The total force should cause a displacement pointing toward $-(\vec{R}_F-\vec{R}_{CM})/\lambda_0$, with $\vec{R}_F$ the new CE of the main fields in consideration, Eqs. (\ref{eq46}). Such factor is shape and mode dependent.

However, the most significant conclusion is about the time range of the expected momentum change. This interval will always be the typical transient times for reaching the stationary regime of the fields. For the most important momentum component, Eq. (\ref{eq47b}), internal thrusts can only last a few microseconds when typical values of $Q$ are considered in the frequency range of operation of copper-made resonators. Using the last result of Eq. (\ref{eqlast}) in Section (\ref{momentumvariation}), a transient interval of 2.94 $\mu$s corresponds to a frequency of 340 KHz for an input pulsed source. The onset of a single impulse lasting microseconds might be completely outweighed by the presence of other forces, including those of the same magnitude, but lasting much longer. Consequently, one concludes that any displacement of the cavity body due to an interplay of partial momentum components within the system is unobservable under these conditions. This is specially true considering the existence of other neglected perturbations as discussed in \cite{white2017} and \cite{tajmarspace}. 

As a last consequence of what has been presented here, a cavity body initially at rest (kept fixed) and containing a stationary net EM momentum will be set in motion when released and its generator is turned off. Momentum conservation demands the initial EM momentum present in the fields (in Eq. (4), $\vec{P}(0)\neq 0$) be converted into mechanical motion.  However, in this case, the initial system cannot be considered closed because an external force is necessary to balance the mechanical reaction during the transient of momentum storage.  Therefore, the condition of a stationary CM is not obeyed. Such possibility of momentum exchange of an open system should be explored in the future. 

\section{Funding and support}

The author received no financial support for this research, authorship, and/or publication of this article.

\section{Appendix}

\subsection{Field equality Eq. (\ref{eq7})}\label{eq07derivation}
Writing the LHS of Eq. (\ref{eq7}) in Einstein's index notation
\begin{equation}\nonumber
\int \vec{r}(\vec{J}\cdot\vec{E})d^3r \longrightarrow \mathfrak{I}=\int r_l J_k E_k d^3r,
\end{equation}
one applies Maxwell's equations
\begin{equation}\nonumber
\begin{split}
&\partial_t B_k+\epsilon_{ijk}\partial_i E_j=0,\\
-&\partial_t D_k+\epsilon_{ijk}\partial_i H_j=J_k,
\end{split}
\end{equation}
where $\partial_t$ is the time derivative, $\partial_i$ is the divergence operator, $\epsilon_{ijk}$ is the Levi-Civita symbol, together with the constitutive relations $D_k=\varepsilon E_k$ and $B_k=\mu H_k$, and 
\begin{equation}\nonumber
\epsilon_{ijk}\partial_k(D_iB_j)=\epsilon_{ijk}B_k\partial_id_j-\epsilon_{ijk}D_k\partial_iB_j,
\end{equation}
to obtain
\begin{equation}\nonumber
\begin{split}
\mathfrak{I}&=\int r_l[E_k(-\partial_t D_k+\epsilon_{ijk}\partial_i H_j)]d^3r\\
&=\int r_l[-E_k\partial_t D_k+c^2\epsilon_{ijk} D_k\partial_i B_j]d^3r\\
&=\int r_l[-E_k\partial D_k-H_j\partial B_k-c^2\epsilon_{ijk}\partial_k(D_iB_j)]d^3r.
\end{split}
\end{equation}
Calling
\begin{equation}\nonumber
u=\frac{1}{2}(E_kD_k+H_kB_k),
\end{equation}
the last result can be written as
\begin{equation}\nonumber
\mathfrak{I}=\int r_l[-\partial_t u-c^2\epsilon_{ijk}\partial_k(D_iB_j)]d^3r.
\end{equation}
But
\begin{equation}\nonumber
\partial_k (r_l\epsilon_{ijk}D_iB_j)=\epsilon_{ijk}D_iB_j+r_l\epsilon_{ijk}\partial_k(D_iB_j)
\end{equation}
so that
\begin{equation}\label{Appeq01}
\mathfrak{I}=-\int r_l\partial_t d^3r +c^2\int \epsilon_{ijk}D_iB_jd^3r-c^2\int \partial_k (r_l\epsilon_{ijk}D_iB_j)d^3r.
\end{equation}
\begin{figure}[ht]
\centering
 \includegraphics[width=70mm]{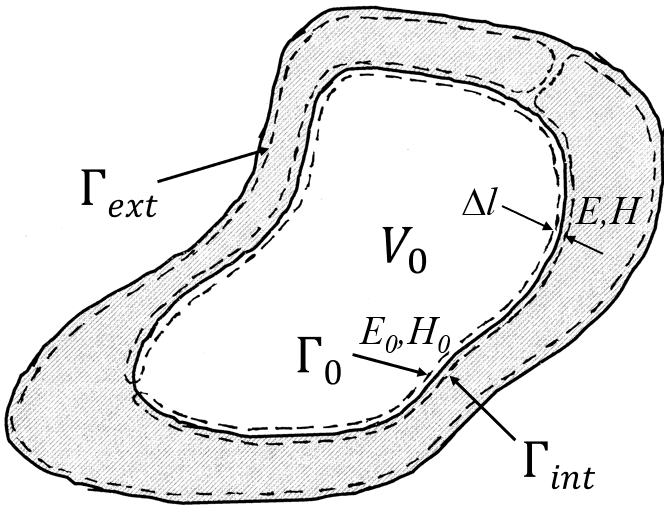}
  \caption{Integration surfaces for solving Eq. (\ref{Appeq01})}
  \label{fig:01app} 
\end{figure}
The last integral in the Eq. (\ref{Appeq01}) is equal to zero as determined  by the Divergence theorem and the problem's physical domain. In fact, this integrals is written as
\begin{equation}\nonumber
\begin{split}
c^2\int \partial_k (r_l\epsilon_{ijk}D_iB_j)d^3r &=\int \partial_k (r_l\epsilon_{ijk}E_iH_j)d^3r\\
&=\oint r_l\epsilon_{ijk}E_iH_j n_k da,
\end{split}
\end{equation}
with $da$ an element of area and $n_k$ an unitary vector outward to the integration surface (Figure \ref{fig:01app}). The integral on the external surface $\Gamma_{ext}$ is negligible because the cavity is so thick that the evanescent fields in $\Gamma$ are close to zero on $\Gamma_{ext}$. Finally, the integrals on the remaining inner surfaces are zero as well because one can write
\begin{equation}\nonumber
\lim_{\Delta l\rightarrow 0}\int_{\Gamma_{int}+\Gamma_0} \vec{r}(\vec{E}\times\vec{H})\cdot \hat{n}da=\lim_{\Delta l\rightarrow 0}\int_{\Gamma_{int}}\vec{r}E_0\cdot\vec{K}da=0
\end{equation} 
with $\Delta l$ the distance between the surfaces $\Sigma_{int}$ and $\Sigma_0$ (in $V_0$, see Figure \ref{fig:01app}), and $\vec{K}=\hat{n}\times(\vec{H}_0-\vec{H})$ is the current density close to $\Gamma_{int}$. Because the fields are finite \cite{stratton2007electromagnetic}(the cavity has a limited conductivity), the resulting integrals goes to zero as the two inner integration surfaces approach the cavity internal surface. 

Therefore,
\begin{equation}
\int \vec{r}(\vec{J}\cdot\vec{E})d^3r=c^2\int \vec{D}\times\vec{B}\,d^3r-\int \vec{r}\frac{\partial u}{\partial t}\,d^3r.
\end{equation} 

\subsection{Average functions for the resonance condition, Eq. (\ref{eqress})}\label{apresfunc}

The average function to be calculated
\begin{equation}
\bigg\langle\varepsilon(t)^2\bigg\rangle =\frac{1}{\tau\sqrt{2\pi}}\int_{-\infty}^{\infty}e^{-(t'-t)^2/2\tau}\varphi^2_{\omega\rightarrow\omega_0}(t')dt',
\end{equation}
with $\varphi_{\omega\rightarrow\omega_0}(t)$ given by Eq. (\ref{eqress}), is such that 
\begin{equation}
\bigg\langle\frac{d}{dt}\varepsilon(t)^2\bigg\rangle =\frac{d}{dt}\bigg\langle\varepsilon(t)^2\bigg\rangle.
\end{equation}

The required integrals are given explicitly by the following expression 
\begin{equation}\label{eq24_0}
\begin{split}
\bigg\langle\varepsilon(t)^2\bigg\rangle &=\left(\frac{f_0^2\omega_0^2}{\Gamma_0^2}\right)\Bigg\{\frac{1}{2}\left(1-e^{-2\omega_0^2\tau^2}\cos 2\omega_0 t\right)+\Bigg.\\
&\frac{1}{2\lambda^2}\left[e^{2(1-\lambda^2)\omega_0^2\tau^2}-e^{-\lambda(1+\lambda)\omega_0^2\tau^2}\cos\omega_0(1+\lambda)(t-\Gamma_0\tau^2/2)\right]e^{-\Gamma_0 t}-\\
&\Bigg.\frac{1}{\lambda}\left[e^{\lambda (1-\lambda^2)\omega_0^2\tau^2}\cos\omega_0(1-\lambda)(t-\Gamma_0\tau^2/2)-e^{-\lambda(1+\lambda)\omega_0^2\tau^2}\cos\omega(1+\lambda)(t-\Gamma_0\tau^2/2)\right]e^{-\Gamma_0 t/2}\Bigg\},
\end{split}
\end{equation}
and 
\begin{equation}\label{eq24}
\begin{split}
\bigg\langle\frac{d}{dt}\varepsilon(t)^2\bigg\rangle &=\left(\frac{f_0^2\omega_0^2}{\Gamma_0^2}\right)\Bigg\{ \omega_0e^{-2\omega_0^2\tau^2}\sin 2\omega_0t\Bigg.\quad -\\
&\omega_0e^{-\Gamma_0 t/2}\left[\left(1+\frac{1}{\lambda}\right)e^{-\lambda (1+\lambda)\omega_0^2\tau^2}\sin\omega_0(1+\lambda)(t-\Gamma_0\tau^2/2)\right. -\\
&\;\quad\quad\quad\quad \left.\left(1-\frac{1}{\lambda}\right)e^{\lambda (1-\lambda)\omega_0^2\tau^2}\sin\omega_0(1-\lambda)(t-\Gamma_0\tau^2/2)\right]+\\
&\left(\frac{\Gamma_0}{2\lambda}\right)e^{-\Gamma_0 t/2}\Bigg[e^{\lambda (1-\lambda)\omega_0^2\tau^2}\cos\omega_0(1-\lambda)(t-\Gamma_0\tau^2/2)\Bigg.\\
& \quad\quad\quad\quad\quad -e^{-\lambda (1+\lambda)\omega_0^2\tau^2}\cos\omega_0(1+\lambda)(t-\Gamma_0\tau^2/2)\Bigg] +\\
& \left(\frac{\omega_0}{\lambda}\right) e^{-2 (2\lambda^2-1)\omega_0^2\tau^2-\Gamma_0 t}\sin 2\lambda\omega_0(t-\Gamma_0\tau^2) -\\
&\Bigg.\left(\frac{\Gamma_0}{2\lambda^2}\right)e^{-\Gamma_0 t}\Bigg[ e^{2(1-\lambda^2)\omega_0^2\tau^2}- e^{-2(2\lambda^2-1)\omega_0^2\tau^2}\cos 2\lambda\omega_0(t-\Gamma_0\tau^2) \Bigg]\Bigg\}.
\end{split}
\end{equation}
As $\Gamma_0/\omega_0 \ll 1$, $\omega_0\tau \gg 1$, $\lambda \approx 1$ and $1-\lambda \approx (1/2)(\Gamma_0/2\omega_0)^2$, the only remaining terms in the integral allow us to further simplify
\begin{equation}
\begin{split}
\bigg\langle\varepsilon(t)^2\bigg\rangle &\approx \left(\frac{f_0^2\omega_0^2}{2\Gamma_0^2}\right)\Bigg\{1+e^{-\Gamma_0 t}-2e^{-\Gamma_0 t2/2+\Gamma_0^2\tau^2/8}\cos\left[\frac{\Gamma_0^2}{8\omega_0^2}(t-\Gamma_0\tau^2/2)\right]\Bigg\},\\
\bigg\langle\frac{d}{dt}\varepsilon(t)^2\bigg\rangle &\approx\left(\frac{f_0^2\omega_0^2}{2\Gamma_0}\right)e^{-\Gamma_0 t/2+\Gamma_0^2\tau^2/8}\left[\cos\frac{\Gamma_0^2}{8\omega_0}\left(t-\frac{\Gamma_0\tau^2}{2}\right)-e^{-\Gamma_0 t/2+3\Gamma_0^2\tau^2/8}\right].
\end{split}
\end{equation}
Eqs. (\ref{eq19}) are based on these last equations, by further expanding $\cos x\approx 1-x^2/2$ in the limit $\Gamma_0\rightarrow 0$, and by neglecting $\Gamma_0 \tau^2\approx 0$.

\subsection{Center of energy generated by Eq. (\ref{eq31})}\label{centerenergy}

Defining
\begin{equation}\label{eq39}
\begin{split}
\mathcal{I}_{C1}(\xi,p)^{(m,n)}&=\int_{0}^{1}(1+\xi u)^m\cos^n p\pi u \;du, \;\;\mathcal{I}_{C2}(\xi,p)^{(m,n)}=\int_{0}^{1}u(1+\xi u)^m\cos^n p\pi u \;du, \\
\mathcal{I}_{S1}(\xi,p)^{(m,n)}&=\int_{0}^{1}(1+\xi u)^m\sin^n p\pi u \;du, \;\;\mathcal{I}_{S2}(\xi,p)^{(m,n)}=\int_{0}^{1}u(1+\xi u)^m\sin^n p\pi u \;du,
\end{split}
\end{equation}
with $m$ and $n$ positive integers, the magnetic and electric energy centers are located at ($z_b=0$)
\begin{equation}\label{eq40}
\begin{split}
\vec{R}_{H}&=\vec{z}L\frac{\mathcal{I}_{C2}(p,\xi)^{(4,2)}}{\mathcal{I}_{C1}(p,\xi)^{(4,2)}},\\
\vec{R}_{E}&=\vec{z}L\left[\frac{3\psi_{\varphi}^2\mathcal{I}_{S2}(p,\xi)^{(4,2)}+2\psi_0^2\rho_1^2\mathcal{I}_{C2}(p,\xi)^{(6,2)}}{3\psi_{\varphi}^2\mathcal{I}_{S1}(p,\xi)^{(4,2)}+2\psi_0^2\rho_1^2\mathcal{I}_{C1}(p,\xi)^{(6,2)}}\right]\\
\vec{R}_{H_{\Gamma}}&=\vec{z}L\left[\frac{\rho_1(1+\xi)^4+4\sqrt{\xi^2\rho_1^2+1}\mathcal {I}_{C2}(\xi,p)^{(3,2)}}{\rho_1[1+(1+\xi)^4]+4\sqrt{\xi^2\rho_1^2+1}\mathcal {I}_{C1}(\xi,p)^{(3,2)}}\right]
\end{split}
\end{equation}
with $\psi_{\varphi}=2p\pi \psi_0/\gamma^2 L^2$.

\bibliography{sample}

\begin{thebibliography}{10}

\bibitem{sherburne1953momentum}
R.~K. Sherburne and W.~L. Weeks.
\newblock Momentum thrust of a rocket.
\newblock {\em American Journal of Physics}, 21(2):139--140, 1953.

\bibitem{wickman1981technology}
J.~Wickman.
\newblock Technology assessment of photon propulsion-how close are we.
\newblock In {\em 17th Joint Propulsion Conference}, page 1532, 1981.

\bibitem{bae2012prospective}
Y.~K. Bae.
\newblock Prospective of photon propulsion for interstellar flight.
\newblock {\em Physics Procedia}, 38:253--279, 2012.

\bibitem{marx1966interstellar}
G.~Marx.
\newblock Interstellar vehicle propelled by terrestrial laser beam.
\newblock {\em Nature}, 211(5044):22, 1966.

\bibitem{white2017}
H.~White, P.~March, J.~Lawrence, J.~Vera, A.~Sylvester, D.~Brady, and
  P.~Bailey.
\newblock Measurement of impulsive thrust from a closed radio-frequency cavity
  in vacuum.
\newblock {\em Journal of Propulsion and Power}, 33(4):830--841, 2017.

\bibitem{duif2017improved}
C.~P. Duif.
\newblock An improved method to measure microwave induced impulsive forces with
  a torsion balance or weighing scale.
\newblock {\em arXiv preprint arXiv:1706.04999}, 2017.

\bibitem{kossling2019spacedrive}
M.~K{\"o}{\ss}ling, M.~Monette, M.~Weikert, and M.~Tajmar.
\newblock The spacedrive project-thrust balance development and new
  measurements of the mach-effect and emdrive thrusters.
\newblock {\em Acta Astronautica}, 161:139--152, 2019.

\bibitem{mullins2006fly}
J.~Mullins.
\newblock Fly by light.
\newblock {\em New Scientist}, 191(2568):30--34, 2006.

\bibitem{shawyer2015second}
R.~Shawyer.
\newblock Second generation emdrive propulsion applied to ssto launcher and
  interstellar probe.
\newblock {\em Acta Astronautica}, 116:166--174, 2015.

\bibitem{juan2013prediction}
Yang Juan, Wang Yu-Quan, Ma~Yan-Jie, Li~Peng-Fei, Yang Le, Wang Yang, and
  He~Guo-Qiang.
\newblock Prediction and experimental measurement of the electromagnetic thrust
  generated by a microwave thruster system.
\newblock {\em Chinese Physics B}, 22(5):050301, 2013.

\bibitem{mcculloh2017}
M.~E. McCulloh.
\newblock Testing quantised inertia on emdrives with dielectrics.
\newblock {\em Europhysics Letters}, 118(3), 2017.

\bibitem{fetta2014numerical}
G.~P. Fetta.
\newblock Numerical and experimental results for a novel propulsion technology
  requiring no on-board propellant.
\newblock In {\em 50th AIAA/ASME/SAE/ASEE Joint Propulsion Conference}, page
  3853, 2014.

\bibitem{grahn2016exhaust}
P.~Grahn, A.~Annila, and E.~Kolehmainen.
\newblock On the exhaust of electromagnetic drive.
\newblock {\em AIP Advances}, 6(6):065205, 2016.

\bibitem{tajmarspace}
M.~Tajmar, O.~Neunzig, and M.~Weikert.
\newblock Hight-accuracy thrust measurements of the emdrive and elimination of
  false-positive effects.
\newblock {\em Space Propulsion 2020+1}, SP2020-268, 2021.

\bibitem{einstein1906prinzip}
A.~Einstein.
\newblock Das prinzip von der erhaltung der schwerpunktsbewegung und die
  tr{\"a}gheit der energie.
\newblock {\em Annalen der Physik}, 325(8):627--633, 1906.

\bibitem{keller1942newton}
J.~M. Keller.
\newblock Newton's third law and electrodynamics.
\newblock {\em American Journal of Physics}, 10(6):302--307, 1942.

\bibitem{zangwill2013modern}
A.~Zangwill.
\newblock {\em Modern electrodynamics}.
\newblock Cambridge University Press, 2013.

\bibitem{coleman1968origin}
S.~Coleman and J.~H. Van~Vleck.
\newblock Origin of ``hidden momentum forces'' on magnets.
\newblock {\em Physical Review}, 171(5):1370, 1968.

\bibitem{calkin1971linear}
M.~G. Calkin.
\newblock Linear momentum of the source of a static electromagnetic field.
\newblock {\em American Journal of Physics}, 39(5):513--516, 1971.

\bibitem{pugh1967physical}
E.~M. Pugh and G.~E. Pugh.
\newblock Physical significance of the poynting vector in static fields.
\newblock {\em American Journal of Physics}, 35(2):153--156, 1967.

\bibitem{graham1980observation}
G.~M. Graham and D.~G. Lahoz.
\newblock Observation of static electromagnetic angular momentum in vacua.
\newblock {\em Nature}, 285(5761):154--155, 1980.

\bibitem{stratton2007electromagnetic}
J.~A. Stratton.
\newblock {\em Electromagnetic theory}, volume~33.
\newblock John Wiley \& Sons, 2007.

\bibitem{boyer2005illustrations}
T.~H. Boyer.
\newblock Illustrations of the relativistic conservation law for the center of
  energy.
\newblock {\em American journal of physics}, 73(10):953--961, 2005.

\bibitem{moller1972theory}
C.~M{\o}ller.
\newblock {\em The theory of relativity}.
\newblock Ofxord at the Claredon Press, 1972.

\bibitem{penfield1966electrodynamics}
P.~L. Penfield~Jr and H.~A. Haus.
\newblock {\em Electrodynamics of moving media}.
\newblock MIT, Cambridge, MA, USA, 1967.

\bibitem{collin1990field}
R.~E. Collin.
\newblock {\em Field theory of guided waves}, volume~5.
\newblock John Wiley \& Sons, New York, 1990.

\bibitem{ilinskiui1993propagation}
A.~S. Ilyinsky, G.~Y. Slepyan, and A.~Y. Slepyan.
\newblock {\em Propagation, scattering and dissipation of electromagnetic
  waves}.
\newblock Number~36. Peter Peregrinus Ltd, 1993.

\bibitem{karlsson2014microwave}
A.~Karlsson and G.~Kristensson.
\newblock Microwave theory.
\newblock {\em Tryckeriet i E-huset, Lund University, Lund}, 2014.

\bibitem{pozar2011microwave}
D.~M. Pozar.
\newblock {\em Microwave engineering}.
\newblock John Wiley \& Sons, 2011.

\bibitem{jackson1999classical}
J.~D. Jackson.
\newblock {\em Classical electrodynamics}.
\newblock John Wiley \& Sons., 1999.

\bibitem{yang2016thrust}
J.~Yang, X.~Liu, Y.~Wang, M.~Tang, L.~Luo, Y.~Jin, and Z.~Ning.
\newblock Thrust measurement of an independent microwave thruster propulsion
  device with three-wire torsion pendulum thrust measurement system.
\newblock {\em Journal of Propulsion Technology}, 37(2):362--371, 2016.

\end{thebibliography}

\end{document}